\documentstyle{article}
\begin{document}

\title {\bf  Strong BBP couplings for the charmed baryons}
\author{
R. C. Verma$^{*}$\\
Theoretical Physics Institute,
Department of Physics\\
University of Alberta, Edmonton, Alberta, {\bf Canada}\\
M. P. Khanna\\
Centre for Advanced Study in Physics, Department of Physics, \\
Panjab University, Chandigarh -160 014 {\bf India}
}
\maketitle
\begin{abstract}
\large
According to the Coleman-Glashow null theorem if all the symmetry
breaking effects
belong to the same regular representation (octet in the case of SU(3)
and {\bf 15}
in the case of SU(4)) and are generated in a tadpole type mechanism,
the strangeness
changing (charm changing)  weak transitions generated through the
S$_{6}$ (S$_{9}$) tadpole
must vanish. Exploiting this null result, we find relations between
the BBP
coupling constants which allow us to write the coupling constants in
terms
of two parameters and baryon masses. Fixing these two parameters (
$g_{NN\pi}$
and $g_{\Lambda \Sigma \pi}$) from experiments, we estimate   the
remaining
coupling constants.
\end{abstract}
\vskip 1.0 cm
PACS Index:  11.30Hv, 13.30 Eg, 11.40 Ha \\
\vskip 4.0 cm
$^{*}$  On Leave from Centre for Advanced Study in Physics,
Department of Physics, \\
Panjab University, Chandigarh -160 014 {\bf India}

\newpage
\Large
\section{Introduction}
\par
{}From the strong coupling constants extracted from data, it seems very
well
established that SU(3) invariance does not work for the BBP strong
couplings,
$g_{B_{j}B_{k}P_{i}}$, the Yukawa coupling of the pseudoscalar mesons
with
baryons. Some form of symmetry breaking must be invoked in order to
account
for the observed couplings.
\par
In a simple method [1,2] the SU(3) symmetry breaking at the  BBP
vertex could be accounted for by exploiting the Coleman-Glashow null
theorem  [3]
for the tadpole type symmetry breaking. For the symmetry breaking
effects,
the medium strong (responsible for the mass diferences in the
internal symmetry
multiplets), the electromagnetic and the weak effects, transforming
as the members
of the same octet, the strangeness changing scalar tadpole S$_{6}$
can be
rotated away by a unitary transformation. Therefore, the strangeness
changing
transitions generated through S$_{6}$ must vanish. This is the
Coleman-Glashow
theorem. The current algebra and the Partially Conserved Axial Vector
Current
(PCAC) hypothesis have been used to write expression for the
amplitudes of
non-leptonic decays of hyperons in terms of the strong BBP coupling
constants
and baryon masses by Marshak et al. [4] and others [5] in the quark
density model. The
Coleman-Glashow theorem requires that these expressions for the
parity conserving
modes must all be equal to zero. The decays can, however, arise
through current  $\otimes$
current form of the weak interactions, which can be fairly
satisfactorily
explained on the basis of the standard model. Marshak et al. [4] have
explicitly shown that these amplitude expressions due to tadpole
indeed go
to zero if all couplings and masses are SU(3) invariant. The null
result has
also been shown [6] to hold when the SU(3) symmetry for the couplings
and for
the masses is assumed broken via the tadpole mechanism. The
requirement
that each of the weak non-leptonic decay amplitude generated through
the
S$_{6}$ tadpole must be zero, then gives relations among the symmetry
broken
coupling constants and baryon masses. For the eight BBP coupling
constants
involving pions and kaons, six relations are obtained [1,2]. Thus the
eight
coupling constants are determined in terms of only two parameters, no
more
than the SU(3) symmetric case. The SU(3) symmetry broken BBP
couplings
so obtained are in good agrrement with the available experimental
numbers [7-10].
\par
The null result proved initailly for the SU(3) case is valid for the
SU(4)
case also if the symmetry breaking effects belong to the regular
representation
{\bf15.} The current $\otimes$ current weak Hamiltonian responsible
for the hadronic
weak decays of charmed baryons belongs to the  {\bf 20}$^{\prime
\prime}$ and

{\bf 84} representations of the SU(4). The tadpole part of the weak
Hamiltonian will belong to
the {\bf 15} representation. And it is this piece of the Hamiltonian
which will give
zero amplitude for the decays. The considerations can thus be
generalised to
the charm sector and relations for the strong couplings obtained.
This was done
[2] for some  of the charm baryon coupling constants.

\par
In this work, we do a detailed analysis
and calculate all the strong coupling constants of charm baryons
involving $\pi$, K, D  and $D_{s} $  mesons by considering  $\pi$ and
K  emitting parity conserving
decays whether kinematically allowed or not. Now that the hadronic
properties
of heavier baryons have become a real possibilty, the strong coupling
constants
involving charmed baryons are of immediate interest for studying
them, e.g., for studying weak hadronic decays of charm baryons [11,12
]. We develop our formalism in section (2) and list our results in
section  (3) giving relations among various strong couplings.
Numerical estimates are given in section (4).

\section{Non-leptonic Baryon Decays}
\par
The matrix element for the baryon decay process
$$ B_{j} \rightarrow B_{k} + P_{i} \eqno(1) $$
can be expressed as
$$ M = - \langle B_{k} P_{i}|H_{W} | B_{j}\rangle = \bar {u}_{B_{k}}
[i A + B \gamma_{5} ] u_{B_{j}} \eqno(2) $$
where $j,~ k $ represent internal symmetry (SU(3), SU(4)) indices for
the
initial and final baryons and $i$ is the index for the emitted
pseudoscalar
meson. A and B are the  pv (parity-violating)  and the  pc
(parity-conserving)
decay amplitudes respectively. This three hadron matrix element can
be
reduced [4,5] to the baryon-baryon transition matrix element of
H$_{W}$ by the
application of current algebra along with the PCAC hypothesis as
$$
M = \frac {i}{f_{P}} \langle B_{k} |[Q^{5}_{i},~ H_{W}]|B_{j} \rangle
+ M_{P} \eqno (3) $$
where $Q_{i}^{5}$ is the axial generator associated with the meson
P$_{i}$
and f$_{P}$ is the corresponding pseudoscalar meson decay constant.
The
contribution from the pole diagram is contained in M$_{P}$ .

\subsection{Quark density model}
\par
The tadpole piece of the weak Hamiltonian in SU(3), for example, is
written as
$$
H_{W}^{\Delta S = 1} = G_{pv} P_{7} + G_{pc}S_{6} \eqno(4) $$
where S$_{6}$ and P$_{7}$ are the scalar and pseudoscalar quark
densities transforming as $\lambda_{6}$ and $\lambda_{7}$ components
of an
SU(3) octet respectively.
\par
The symmetry breaking in mass is described by
$$ M = M_{0} + \delta M_{s} S_{8} + \delta M_{em} S_{3} \eqno(5)$$
The assumption that $S_{8}$ the medium strong, $S_{3}$ the
electromagnetic
and $S_{6}$ the weak symmetry breaking tadpoles belong to the same
octet is
called the universality hypothesis for the spurions. However, one can
always find an SU(3) transformation  such that the strangeness
changing scalar tadpole $S_{6}$ can be rotated away leaving behind
the medium strong and electromagnetic pieces  $S_{8}$ and $S_{3}$
respectively  as the diagonal eighth and third components of the
SU(3) octet. The strangeness changing effects generated through
$S_{6}$  must, therefore, vanish. Within the framework of current
algebra this can be seen easily [4,5].
\par
 For the pc-weak Hamiltonian (4), the general expression (3) gives
for the
p-wave amplitudes
$$ \langle B^{\prime}_{k} P_{i}|H_{W}^{PC} | B_{j}\rangle =
- \frac {\sqrt 2}{f_{P}} d_{ism} \langle B_{k} | P_{m}| B_{j} \rangle
$$

$$
+ \frac {a_{jm} g_{mki}}{M_{j} - M_{m}} \frac {M_{k} + M_{j}}{M_{k} +
M_{m}}
+ \frac { g_{jni}a_{nk}}{M_{k} - M_{n}} \frac {M_{j} + M_{k}}{M_{j} +
M_{n}}\eqno(6)$$
where $m,~ n$ are indices for the intermediate baryons in the s-
channel
and u-channel respectively. $a_{jm}$ is the PC weak transition
amplitude of
$B_{j} \rightarrow B_{m}$. $M_{j}$ is the mass of $B_{j}$, $s$ is
unitary index
for the weak spurion with fixed value 6 for the strangeness changing
weak baryon decays. Using the universality of the spurions and the
SU(3) symmetry for the strong
couplings and the baryon masses we see [3] that the decay amplitudes
in equation (6)
are all zero if we take $ \delta M_{s} = - \frac {2} {\sqrt 3}
f_{\pi} $.
\par
It is straightforward to generalize this formalism to the charm
sector. Quark-density weak Hamiltonian for charm changing ($\Delta C$
= -1, $\Delta S$ = 0) mode is given by
$$ H_{W}^{ \Delta C = -1, \Delta S = 0} = G_{pv} P_{10} + G_{pc}S_{9}
\eqno(7)$$
and the mass breaking is given by

$$ M = M_{0} + \delta M_{c} S_{15} + \delta M_{s} S_{8} + \delta
M_{em} S_{3} \eqno(8)$$
Here, mass breaking terms $S_{15}$, $S_{8}$ and  $S_{3}$ and the weak
symmetry breaking tadpole $S_{9}$ are assumed to belong to the same
{\bf 15}  representation of the SU(4). Now the index  s in eq.(6)
would become 9 for the charm changing and strangeness conserving
decays.
\section{Relations among  strong couplings}

\par
By considering various p-wave decay modes in the quark density model
and setting them
equal to zero, we obtain relations containing strong coupling
constants, meson decay constants and  baryon masses. We list the
results according to the decay modes involved.

\subsubsection{$\pi$ Couplings}
\par
There are ten SU(2) invariant couplings involving pions. The four
couplings
$NN \pi$, $\Lambda \Sigma \pi$, $\Sigma \Sigma \pi$, $\Xi \Xi \pi$
exist
for the SU(3) octet (C=0) baryons  ( $ N, \Lambda, \Sigma, \Xi $)
whereas the remaining  $\Sigma_{c} \Lambda_{c} \pi$, $\Sigma_{c}
\Sigma_{c} \pi$
$\Xi^{\prime}_{c} \Xi^{\prime}_{c} \pi$, $\Xi_{c} \Xi^{\prime}_{c}
\pi$, $\Xi_{c} \Xi_{c} \pi$ for the {\bf 3}$^{*}$-$( \Lambda_{c},
\Xi^{\prime}_{c} )$ and {\bf 6}-$(\Sigma_{c}, \Xi_{c}, \Omega_{c} )$
of charm baryons (C=1) and $\Xi_{cc} \Xi_{cc} \pi$ for {\bf 3}-$(
\Xi_{cc} , \Omega_{cc} )$  of (C=2) charm baryons. \\
\subsubsection{K couplings}
\par
The K couplings are $\Lambda N K,~ \Sigma N K, ~ \Xi \Lambda K, \Xi
\Sigma K$ for the
octet baryons corresponding to C=0,
$\Xi_{c}^{\prime} \Lambda _{c} K$,  $\Xi_{c}^{\prime} \Sigma _{c} K$,
$\Xi_{c}\Lambda _{c} K$, $\Xi_{c} \Sigma _{c} K$, $\Omega_{c} \Xi_{c}
K$ and    $\Omega_{c} \Xi_{c}^{\prime} K$   from representations {\bf
3}$^{*}$
and {\bf 6} corresponding to C= 1 and $\Omega_{cc} \Xi_{cc} K$ from
{\bf 3} corresponding to C=2 baryons. The total number of coupling
constants here is eleven.

\subsection{ Octet baryon strong couplings}
\par
{}From all the baryon decays in $\Delta C=0,~ \Delta S= 1$ mode,
whether kinematically allowed or not, we get six independent
relations for the coupling constants. These relations allow us to
write all the
eight pion and kaon couplings for the octet baryons in terms of two
parameters
for which we choose $g_{NN\pi}$ and $g_{\Lambda \Sigma \pi}$. \\

We consider the six decays $\Sigma^{+} \rightarrow n \pi^{+}$,~
$\Xi^{0} \rightarrow
\Sigma^{-} \pi^{+}$,~ $\Lambda \rightarrow p \pi^{-}$,~ $\Sigma^{-}
\rightarrow n \pi^{-}$,~
$\Xi^{0} \rightarrow p K^{-}$ and $\Xi^{-} \rightarrow n K^{-}$ and
compute their
amplitudes by current algebra and soft pion technique [4]  taking the
quark density model
for the weak Hamiltonian. Setting each of these amplitudes to zero,
we get:

\subsubsection{ $\Delta C=0,~ \Delta S= 1 : \pi$-meson emitting
decays of the hyperons}
\par
The decays $\Sigma^{+} \rightarrow n \pi^{+}$ and $\Xi^{0}
\rightarrow \Sigma^{-}\pi^{+}$
yield two relations:
$$
G_{\Sigma \Sigma  \pi }
=  -2 G_{N N \pi }    +  {\sqrt 3} G_{\Lambda \Sigma \pi }  \eqno(9)
$$
$$
 G_{\Xi \Xi  \pi }
=  G_{N N \pi }  - {\sqrt 3} G_{\Lambda \Sigma \pi }   \eqno(10)
$$

Here, we have used  the following definition:

$$ G_{B_{j} B_{k} \pi} =  g_{B_{j} B_{k}\pi}\frac {2M_{N}}{M_{B_{j}}
+ M_{B_{k}}} $$

The decays $\Lambda \rightarrow p \pi^{-}$ and  $\Sigma^{-}
\rightarrow n \pi^{-}$
give
$$ G_{\Lambda N K} = \frac {{\sqrt 2}f_{\pi}}{\delta M_{s}} ( {\sqrt
2} G_{NN\pi}
- \frac {2} {\sqrt 6} G_{\Lambda \Sigma \pi}) \eqno(11)$$
$$ G_{\Sigma N K} =  \frac {{\sqrt 2}f_{\pi}}{\delta M_{s}} ( \frac
{2}{\sqrt 6} G_{NN\pi}
-  {\sqrt 2}  G_{\Lambda \Sigma \pi}) \eqno(12)$$

\subsubsection{ $\Delta C= 0,~ \Delta S = 1 : K$-meson emitting
decays of the hyperons}
The decays $\Xi^{0} \rightarrow p K^{-}$ and $\Xi^{-} \rightarrow n
K^{-}$
give rise to the following relations:
$$ G_{\Xi \Sigma K} = {1 \over 2} ( - {\sqrt 3} G_{\Lambda N K} +
G_{\Sigma N K})
\eqno(13)$$
$$ G_{\Xi \Lambda K} = - {1 \over 2} (  G_{\Lambda N K} + {\sqrt
3}G_{\Sigma N K})
\eqno(14)$$

Substitution of Eqs. (11) and (12) into the equations (13) and (14)
allows us to
express $g_{\Xi \Sigma K}$ and $g_{\Xi \Lambda K}$ in terms of
$g_{NN\pi}$ and $g_{\Lambda \Sigma \pi}$,
$$ G_{\Xi \Lambda K} =  \frac {{\sqrt 2}f_{\pi}}{\delta M_{s}} (-
{\sqrt 2} G_{NN\pi}
+ \frac {4} {\sqrt 6} G_{\Lambda \Sigma \pi}) \eqno(15)$$
$$ G_{\Xi \Sigma K} = \frac {{\sqrt 2}f_{\pi}}{\delta M_{s}}
(- \frac {2} {\sqrt 6} G_{NN\pi})
 \eqno(16)$$

Equations (9), (10), (11) and (12)
already express the coupling constants $g_{\Sigma \Sigma \pi}$,
$g_{\Xi \Xi \pi}$, $g_{\Lambda N K}$ and
$g_{\Sigma N K}$ in terms of  $g_{NN\pi}$ and $g_{\Lambda \Sigma
\pi}$.

\subsection{  K-meson couplings of charm baryons}
\par
We now consider the K-meson emitting  weak decays of charmed baryons
in the similar manner. $\Delta C= -1,~ \Delta S=0$  decays
$\Xi^{\prime 0}_{c} \rightarrow p K^{-}$,
$\Xi_{c}^{0} \rightarrow p K^{-}$,~
$\Lambda_{c} \rightarrow \Sigma^{0} K^{+}$,~
$\Sigma^{+}_{c} \rightarrow \Lambda K^{+}$   give rise to the
following relations:
$$ G_{\Xi^{\prime}_{c} \Lambda_{c} K} = \
\frac {1}{2 {\sqrt 6}}G_{\Lambda N K} +
\frac {3}{2 {\sqrt 2}}G_{\Sigma N K} \eqno(17)$$

$$ G_{\Xi^{\prime}_{c} \Sigma_{c} K} = \frac {1}{2 {\sqrt
2}}G_{\Lambda N K} -
\frac {3}{2 {\sqrt 6}}G_{\Sigma N K} \eqno(18)$$
$$ G_{\Xi_{c} \Lambda_{c} K} = \frac {1}{2 {\sqrt 2}}G_{\Lambda N K}
-
\frac {3}{2 {\sqrt 6}}G_{\Sigma N K} = G_{\Xi^{\prime}_{c} \Sigma_{c}
K}\eqno(19)$$
$$ G_{\Xi_{c} \Sigma_{c} K} = \frac {3}{2 {\sqrt 6}}G_{\Lambda N K} +
\frac {1}{2 {\sqrt 2}}G_{\Sigma N K} \eqno(20)$$
Similarly the consideration of the $\Xi_{c}/\Xi_{c}^{\prime}
\rightarrow \Xi + K $ modes gives

$$ G_{\Omega_{c} \Xi_{c}^{\prime} K}  = \frac {1}{2} (G_{\Lambda \Xi
K} - {\sqrt 3}
G_{\Sigma \Xi K} )\eqno(21)$$
$$ G_{\Omega_{c} \Xi_{c} K}  = \frac {1}{2} ({\sqrt 3} G_{\Lambda \Xi
K }+
G_{\Sigma \Xi K} )\eqno(22)$$
Further the decay $\Omega_{cc} \rightarrow \Sigma_{c}^{++} K^{-}$
yields
$$ G_{\Omega_{cc} \Xi_{cc} K}  = ({\sqrt 3}
G_{\Xi_{c}^{\prime}\Sigma_{c} K} -
G_{\Xi _{c}\Sigma_{c}  K} )\eqno(23)$$
Note that all the charm baryon K-couplings are related to the octet
baryon K-couplings. Finally, these are expressed in terms of
$g_{\Lambda \Sigma \pi}$ and $g_{N N \pi}$ as follows:
$$ G_{\Xi^{\prime}_{c} \Lambda_{c} K} = \frac {{\sqrt
2}f_{\pi}}{\delta M_{s}}
(\frac {2} {\sqrt 3} G_{NN\pi} - \frac {5} {3} G_{\Lambda \Sigma
\pi}) \eqno(24)$$
$$ - \frac {1} {\sqrt 2} G_{\Omega_{c} \Xi_{c}^{\prime} K} =
G_{\Xi_{c} \Lambda_{c} K} = G_{\Xi^{\prime} \Sigma_{c} K} =  \frac
{{\sqrt 2}f_{\pi}}{\delta M_{s}} (
 \frac {1} {\sqrt 3} G_{\Lambda \Sigma \pi}) \eqno(25)$$
$$ - \frac {1} {\sqrt 2} G_{\Omega_{c} \Xi_{c} K} =  G_{\Xi_{c}
\Sigma_{c} K} =  \frac {{\sqrt 2}f_{\pi}}{\delta M_{s}}
 (\frac {2} {\sqrt 3} G_{NN\pi} - G_{\Lambda \Sigma \pi})\eqno(26)$$
$$ G_{\Omega_{cc} \Xi_{cc} K} =  \frac {{\sqrt 2}f_{\pi}}{\delta
M_{s}}
(- \frac {2} {\sqrt 3} G_{NN\pi} +  2 G_{\Lambda \Sigma \pi}) = -
{\sqrt 2} G_{\Sigma N K}\eqno(27)$$

\subsection{$\pi$ meson couplings of  charmed baryons}
\subsubsection{$\pi$-meson emittimg $\Delta C =-1,~ \Delta S= 0$
decays}
These decays lead to the following relations:
$$
{\sqrt 3} G_{\Sigma_{c} \Lambda _{c} \pi } - G_{\Sigma_{c} \Sigma
_{c}  \pi }
= 2 G_{N N \pi }   \eqno(28)     $$
from the $\Sigma_{c}^{0} \rightarrow p\pi^{-}$,
$$
-2 {\sqrt 3} G_{\Xi_{c}^{\prime} \Xi_{c} ^{\prime} \pi } +2
G_{\Xi_{c} \Xi _{c}^{\prime} \pi }
=  G_{\Lambda \Sigma \pi } + {\sqrt 3} G_{\Sigma \Sigma \pi}
\eqno(29)     $$
from the $\Xi_{c}^{\prime 0} \rightarrow \Sigma^{+} \pi^{-}$,
$$
-2 {\sqrt 3} G_{\Xi_{c}\Xi_{c} ^{\prime} \pi } +2 G_{\Xi_{c} \Xi _{c}
\pi }
=  {\sqrt 3} G_{\Lambda \Sigma \pi } -  G_{\Sigma \Sigma \pi}
\eqno(30)     $$
from the $\Xi_{c}^{0} \rightarrow \Sigma^{+} \pi^{-}$,

$$  {\sqrt 2} G_{\Xi_{cc} \Xi_{cc} \pi} =
 - \frac {1}{\sqrt 2} G_{\Sigma_{c}\Sigma_{c}  \pi }
 -\frac {3}  {\sqrt 6} G_{\Lambda_{c} \Sigma _{c}\pi } \eqno(31)
$$
from the $\Xi_{cc}^{++} \rightarrow \Sigma^{+}_{c} \pi^{+}$.

\subsubsection {$\pi$-meson emitting $\Delta C = 0,~ \Delta S= 1$
decays of charm baryons}
\par
However, the $\Delta C = 0,~ \Delta S= 1$  decay $\Omega^{+}_{cc}
\rightarrow
\Xi_{cc}^{++} \pi^{-}$ leads to
$$ G_{\Omega_{c} \Xi_{cc} K} = - \frac {2}{\sqrt 3} (\frac {\sqrt {2}
f_{\pi}}{\delta M_{s}})
G_{\Xi_{cc} \Xi_{cc} \pi}    \eqno(32)$$
 From the K-couplings $G_{\Omega_{c}\Xi_{cc} K}$ turns out to be
equal to $- {\sqrt 2}G_{\Sigma N K}$.
Substituting the expression of $G_{\Sigma N K}$ in terms of
$G_{NN\pi}$ and
$G_{\Lambda \Sigma \pi}$, we obtain
$$ G_{\Xi_{cc} \Xi_{cc} \pi} = G_{NN\pi} - {\sqrt 3} G_{\Lambda
\Sigma\pi } = G_{\Xi \Xi \pi} \eqno(37)$$
This relation combined with Eqns. (28) and (31) leads to
$$G_{\Lambda_{c} \Sigma_{c} \pi} = G_{\Lambda \Sigma \pi}\eqno(34)$$
$$G_{\Sigma_{c} \Sigma_{c} \pi} = {\sqrt 3} G_{\Lambda \Sigma \pi} -
2G_{NN \pi}
= G_{\Sigma \Sigma \pi}\eqno(35)$$
We have thus been able to write the three charmed baryon couplings
$ \Sigma_{c} \Lambda_{c}\pi$, $\Sigma_{c} \Sigma_{c} \pi$, and
$\Xi_{cc} \Xi_{cc}\pi$
in terms of the two basic parameters $g_{N N \pi}$ and $ g_{\Lambda
\Sigma \pi} $. To express the remaining $\pi$-mesons couplings in
terms of these parameters, we consider now the $\Delta C=0,~ \Delta
S=1 $ decays in $ \Xi^{0}_{c} \rightarrow  \Lambda_{c}
 + \pi^{-} $ modes. This decay involves the mass breakings both due
to
$\delta M_{c}$ and $\delta M_{s}$ unlike the decays considered so far
which involve
either of the two mass breakings. Ignoring the term involving $\delta
M_{s} ~/~ \delta M_{c}$, we get
$$ G_{\Xi_{c} \Xi_{c}^{\prime} \pi } = - \frac {G_{\Lambda \Sigma
\pi}}{2}  \eqno(36)$$
Using this result, the relations (29) and (30) yield the following
sumrules:

$$ G_{\Xi_{c} \Xi_{c} \pi } = - \frac {\sqrt 3}{2}G_{\Lambda \Sigma
\pi} + G_{NN\pi}
\eqno(37)$$

$$ G_{\Xi_{c}^{\prime} \Xi_{c}^{\prime} \pi } = - \frac {5} {2 {\sqrt
3}} G_{\Lambda \Sigma \pi}
+ G_{NN\pi} \eqno(38)$$

\subsection{D-meson couplings of charm baryons}
We have eleven $g_{B B^{\prime} D}$ SU(2) invariant coupling
constants:
The seven
$$ g_{\Lambda_{c} N D},~ g_{\Sigma_{c} N D},~ g_{\Xi_{c}^{\prime}
\Lambda D},
{}~g_{\Xi_{c}^{\prime}\Sigma D},~g_{\Xi_{c} \Lambda D},~g_{\Xi_{c}
\Sigma D},g_{\Omega_{c} \Xi D}$$
correspond to $C=1 \rightarrow C = 0 ~+ ~ D$ mode and the four
$$g_{\Xi_{cc}\Sigma_{c} D},~g_{\Xi_{cc} \Lambda_{c}
D},~g_{\Omega_{cc} \Xi_{c} D},~ g_{\Omega_{cc} \Xi_{c}^{\prime} D} $$
 correspond to $ C=2 ~\rightarrow ~ C + 1 ~+~ D$ mode.

Considering the  $\pi$-meson emitting $\Delta C = -1,~ \Delta S= 0$
decays of  the charm baryons, we obtain the following relations:
$$ G_{\Xi_{c}^{\prime} \Sigma D} =  \frac {{\sqrt 2}f_{\pi}}{ 4
\delta M_{c}}
 (- \frac {3} {\sqrt 6}  G_{\Lambda \Sigma \pi} +  \frac {3} {\sqrt
2} G_{\Sigma \Sigma \pi})
 \eqno(39)$$
from the decay $\Xi^{\prime 0}_{c} \rightarrow \Sigma^{-} \pi^{+} $,
$$ G_{\Xi_{c} \Sigma D} = - \frac {{\sqrt 2}f_{\pi}}{ 4 \delta M_{c}}
 ( \frac {3} {\sqrt 2}  G_{\Lambda \Sigma \pi} +  \frac {3} {\sqrt 6}
G_{\Sigma \Sigma \pi})
 \eqno(40)$$
from the decay $\Xi^{0}_{c} \rightarrow \Sigma^{-} \pi^{+} $,
$$ G_{\Lambda_{c} N D} =  \frac {{\sqrt 2}f_{\pi}}{4\delta M_{c}}
(- 6 G_{NN\pi} + 2 {\sqrt 3} G_{\Lambda_{c} \Sigma_{c} \pi})
\eqno(41)$$
from the decay $\Lambda_{c} \rightarrow n \pi^{+} $,
$$ G_{\Sigma_{c} N D} =  \frac {{\sqrt 2}f_{\pi}}{4\delta M_{c}}
2 {\sqrt 3} ( G_{\Sigma_{c} \Sigma_{c} \pi}+ G_{NN\pi})
\eqno(42)$$
from the decay $\Sigma^{+}_{c} \rightarrow n \pi^{+} $,

$$ G_{\Omega_{c} \Xi D} = \frac {{\sqrt 2}f_{\pi}}{4 \delta M_{c}}
2 {\sqrt 6} G_{\Xi \Xi \pi}
\eqno(43)$$
from the decay $\Omega^{0}_{c} \rightarrow \Xi^{-} \pi^{+}$,
$$ G_{\Xi_{c}^{\prime} \Lambda D} = -  \frac {{\sqrt
2}f_{\pi}}{4\delta M_{c}}
3 {\sqrt 2} ( G_{\Lambda \Sigma \pi} + G_{\Xi_{c}^{\prime} \Xi_{c}
\pi}
+ \frac {1} {\sqrt 3}G_{\Xi_{c}^{\prime} \Xi^{\prime}_{c} \pi} )
\eqno(44)$$
from the decay $\Xi_{c}^{\prime +} \rightarrow \Lambda \pi^{+} $,

$$ G_{\Xi_{c} \Lambda D} =  \frac {{\sqrt 2}f_{\pi}}{4\delta M_{c}}
 {\sqrt 6} ( - G_{\Lambda \Sigma \pi} + G_{\Xi_{c}^{\prime} \Xi_{c}
\pi}
+  {\sqrt 3}G_{\Xi_{c}\Xi_{c} \pi} )
\eqno(45)$$
from the decay $\Xi_{c}^{+} \rightarrow \Lambda \pi^{+} $,
$$ G_{\Xi_{cc} \Lambda_{c} D} = - \frac {{\sqrt 2}f_{\pi}}{4\delta
M_{c}}
 ( 2 {\sqrt 3} G_{\Lambda_{c} \Sigma_{c} \pi} + 6 G_{\Xi_{cc}
\Xi_{cc} \pi}
 )
 \eqno(46)$$
from the decay $\Xi_{cc}^{++} \rightarrow \Lambda_{c}^{+} \pi^{+} $,

$$ G_{\Xi_{cc} \Sigma_{c} D} = \frac {{\sqrt 2}f_{\pi}}{4\delta
M_{c}}
(-3 G_{\Lambda_{c} \Sigma_{c} \pi} + {\sqrt 3} G_{\Sigma_{c}
\Sigma_{c} \pi})
 \eqno(47)$$
from the decay $\Xi_{cc}^{++} \rightarrow \Sigma_{c}^{+} \pi^{+} $,
$$ G_{\Omega_{cc} \Xi_{c} D} =   \frac {{\sqrt 2}f_{\pi}}{4\delta
M_{c}}
 (6 G_{\Xi_{c}^{\prime} \Xi_{c} \pi}
- 2 {\sqrt 3}G_{\Xi_{c} \Xi_{c} \pi} )
\eqno(48)$$
from the decay $\Omega_{cc }^{+} \rightarrow \Xi_{c}^{0} \pi^{+} $,

$$ G_{\Omega_{cc} \Xi_{c}^{\prime} D} =  \frac {{\sqrt
2}f_{\pi}}{4\delta M_{c}}
 (6 G_{\Xi_{c}^{\prime} \Xi_{c}^{\prime} \pi}
- 2 {\sqrt 3}G_{\Xi_{c} \Xi_{c}^{\prime} \pi} )
\eqno(49)$$
from the decay $\Omega_{cc }^{+} \rightarrow \Xi_{c}^{\prime 0}
\pi^{+} $.

Thus, all the D-meson strong couplings are  related to the
$\pi$-meson couplings which in turn have been expressed in terms of
the two basic parameters $g_{\Lambda \Sigma \pi } $  and $g_{ N N
\pi}$.

\subsection{D$_{s}$-meson couplings of charm baryons}
There are seven $g_{B B^{\prime}D_{s}}$ couplings;
$$ g_{\Lambda_{c}\Lambda D_{s}},~ g_{\Sigma_{c}\Sigma D_{s}},~
g_{\Xi_{c}\Xi D_{s}},~ g_{\Xi_{c}^{\prime}\Xi D_{s}}$$
corresponding to $C=1 ~ \rightarrow ~ C = 0 ~+~ D_{s}$ modes, and
$$g_{\Xi_{cc}\Xi_{c} D_{s}},~g_{\Xi_{cc}\Xi_{c}^{\prime}
D_{s}},~g_{\Omega_{cc}\Omega_{c} D_{s}}$$
corresponding to  $C = 2 ~ \rightarrow ~ C = 1~+~ D_{s}$ modes.
Applying our formalism to $K^{+}$ meson emitting $\Delta C = -1,~
\Delta S= 0$ decays, we obtain the following relations:

$$ G_{\Lambda_{c} \Lambda D_{s}} =  \frac {{\sqrt 2}f_{K}}{4 \delta
M_{c}}
2 {\sqrt 2} G_{\Lambda N K } \eqno(50)$$
from the decay $\Lambda_{c}^{+} \rightarrow \Lambda K^{+}$,
$$ G_{\Sigma_{c} \Sigma D_{s}} =  - \frac {{\sqrt 2}f_{K}}{4 \delta
M_{c}}
2 {\sqrt 6} G_{\Sigma N K } \eqno(51)$$
from the decay $\Sigma_{c}^{0} \rightarrow \Sigma^{-} K^{+}$,

$$ G_{\Xi_{c} \Xi D_{s}} = \frac {{\sqrt 2}f_{K}}{4\delta M_{c}}
2{\sqrt 3}   G_{ \Sigma N K}
 \eqno(52)$$
from the decay $\Xi^{0}_{c}  \rightarrow \Xi^{-} K^{+} $,

$$ G_{\Xi_{c}^{\prime} \Xi D_{s}} =  \frac {{\sqrt 2}f_{K}}{4\delta
M_{c}} 2{\sqrt 3}   G_{ \Lambda N K}
 \eqno(53)$$
from the decay $\Xi_{c}^{\prime 0} \rightarrow \Xi^{-} K^{+} $,

$$ G_{\Xi_{cc} \Xi_{c} D_{s}} =\frac {{\sqrt 2}f_{K}}{4\delta M_{c}}
( -  3 G_{ \Lambda N K} - {\sqrt 3} G_{\Sigma N K})
 \eqno(54)$$
from the decay $\Xi_{cc }^{+} \rightarrow \Xi_{c}^{0} K^{+} $,

$$ G_{\Xi_{cc} \Xi_{c}^{\prime} D_{s}} = \frac {{\sqrt
2}f_{K}}{4\delta M_{c}}( -{\sqrt 3} G_{ \Lambda N K} - 3 G_{\Sigma N
K})
 \eqno(55)$$
from the decay $\Xi_{cc }^{+} \rightarrow \Xi_{c}^{\prime 0} K^{+} $,
$$ G_{\Omega_{cc} \Omega_{c} D_{s}}= \frac {{\sqrt 2}f_{K}}{4\delta
M_{c}}  {\sqrt 6}
({\sqrt 3} G_{ \Lambda N K} - G_{\Sigma N K})
 \eqno(56)$$
from the decay $\Omega_{cc }^{+} \rightarrow \Omega_{c}^{0} K^{+} $.
Note that all $g_{B B^{\prime} D_{s}}$ couplings are  related to the
two octet  baryon K-couplings $g_{\Sigma N K}$ and $g_{\Lambda N K}$
which have already  been written in terms of  $g_{N N \pi}$ and
$g_{\Lambda \Sigma \pi}$.

\section{Numerical Work}
To obtain the numerical values of the symmetry broken coupling
constants, we
need the values of $g_{N N \pi}, ~ g_{\Lambda \Sigma \pi}, ~f_{\pi}$,
$f_{K}$, $\delta M_{s}$ and $\delta M_{c}$. We
take
$$ \frac {g^{2}_{N N \pi}}{4 \pi} = 14.6,  \frac {g^{2}_{\Lambda
\Sigma \pi}}{4 \pi} = 11.1 $$
from Pilkuhn et al. [7],

$$ f_{\pi}= 132 MeV,~~ f_{K}= 166 MeV,$$
and the mass breakings

$$ \delta M_{s} = - 160 MeV,~~ \delta M_{c}= -1150 MeV.$$
are determined from the baryon mass differences.  Values of the
coupling constants obtained are tabulated in the third column
of table 1 and they are compared with the SU(4) symmetric values
shown  in the second column.
\section{Discussion}
With the help of Coleman-Glashow null result, we are able to
determine symmetry breaking
at the pseudoscalar BBP vertices without introducing any new
parameters. The
estimated coupling constants are way off from the symmetric values
and are in good agreement with experimental values available at
present.

The experimental values for
$g^{2}_{\Sigma \Sigma \pi } /4 \pi$ from the compilation of Nagels et
al. [9] and for
$g^{2}_{\Lambda N K } / 4 \pi $ and $g^{2}_{\Sigma N K} / 4 \pi$
quoted from Baillon et al. [10] are
respectively $13.4 \pm 2.1$, $20.4 \pm 3.7$ and $1.9 \pm 3.2 $.  The
estimated value
of $g^{2}_{\Lambda_{c} \Sigma_{c} \pi} /4 \pi$ is in good agreement
with the current algebra
calculation [13] and estimated value [14] using Melosh transformation
and PCAC. We find that the strong couplings to charm mesons are
reduced in the presence of the SU(4) flavour symmetry breaking. This
provides a significant test of the formalism considered here.
\par

  With more studies on charmed baryons, the coupling constants are
likely to  find use
in calculations, for example, of charmed baryon-antibaryon bound
states [15],
baryon-exchange forces [16], charmed baryon weak decays [17,18],
where because of
the lack of knowledge on broken coupling constants, invariant
couplings have
been used. For weak hadronic decays [11,12],  use of the symmetry
broken coupling constants affects the pole contributions
significantly.
\section {Acknowledgments}
RCV thanks A.N. Kamal for  providing partial support from a grant
from NSERC, Canada and  the Department of Physics, University of
Alberta for their hospitality. MPK  acknowleges financial assistance
from UGC, New Delhi, India.

\newpage
Table Ia    \\
\begin{center}

{\bf Strong $\pi/ K$-Coupling Constants}
\vskip 0.4 cm
\begin{tabular}{ |c| c| c| c|} \hline
Coupling   &  Symmetric Value   &   Symmetry Broken Value   \\
 $B B^{\prime} P$ & of $g^{2}_{B B^{\prime} P}/4 \pi$ & of $g^{2}_{B
B^{\prime} P}/4 \pi$\\  \hline \hline

$NN\pi$ &  14.60$^{*}$ & 14.60$^{*}$ \\
$\Lambda \Sigma \pi$& 11.10$^{*}$ & 11.10 $^{*}$\\
$\Sigma \Sigma \pi$& 3.50 & 14.03 \\
$\Xi \Xi \pi$ & 3.80 & 1.50\\
$\Lambda_{c} \Sigma_{c} \pi$ & 11.10 & 46.79\\
$\Sigma_{c} \Sigma _{c}\pi$ & 3.50 & 59.39\\
$\Xi_{c} \Xi_{c} \pi$ & 0.88 & 17.17 \\
$\Xi_{c} \Xi^{\prime}_{c} \pi$ & 2.77 &13.59 \\
$\Xi^{\prime}_{c} \Xi^{\prime}_{c} \pi$ & 0.98 & 0.08\\
$ \Xi_{cc} \Xi_{cc}\pi $     & 3.80 & 15.00\\ \hline
$\Lambda N K$ & 10.80 & 18.20 \\
$\Sigma N K$ & 3.80 & 0.90 \\
$\Xi \Lambda K$ & 0.002 & 2.20\\
$\Xi \Sigma K$ & 14.60& 24.69\\
$ \Xi_{c} \Sigma_{c} K$ & 1.75 & 29.02\\
$ \Xi_{c}^{\prime} \Sigma_{c} K$ & 5.55 & 22.91\\
$ \Xi_{c} \Lambda_{c} K$ & 5.55  & 29.92\\
$ \Xi_{c}^{\prime} \Lambda_{c} K$ & 1.95 & 0.10\\
$ \Omega_{c} \Xi_{c}  K$ & 3.50 & 66.77\\
$ \Omega _{c} \Xi_{c}^{\prime}  K$ & 11.10 & 52.97\\
$ \Omega _{cc}\Xi_{cc} K$ & 7.60 & 28.46\\
\hline \hline
\end{tabular}
\end{center}
\newpage
Table Ib    \\
\begin{center}

{\bf Strong $D/D_{s}$-Coupling Constants}
\vskip 0.4 cm
\begin{tabular}{ |c| c| c| c|} \hline

Coupling    &  Symmetric  Value  &   Symmetry Broken Value   \\
 $B B^{\prime} P$ & of $g^{2}_{B B^{\prime} P}/4 \pi$ & of $g^{2}_{B
B^{\prime} P}/4 \pi$\\  \hline \hline

$\Lambda _{c}N D$ &  10.8 & 0.90 \\
$ \Sigma_{c} N D$& 3.80 & 0.05\\
$\Xi_{c}  \Lambda D $& 5.70 & 0.09 \\
$\Xi_{c} \Sigma D$ & 1.90 & 0.03\\
$\Xi_{c}^{\prime} \Lambda D $ & 1.80 & 0.18\\
$\Xi_{c}^{\prime} \Sigma D$ & 5.40 & 0.58\\
$ \Omega _{c}\Xi_{c} D $ & 7.60 & 0.15 \\
$\Xi_{cc} \Sigma_{c} D$ & 29.20 & 7.22\\
$\Xi_{cc} \Lambda_{c} D$ & 0.002 & 0.34\\
$ \Omega_{cc} \Xi_{c}^{\prime} D $     & 0.005 & 0.38\\
$ \Omega_{cc} \Xi_{c} D $     & 14.60 & 4.03\\ \hline
$\Lambda_{c} \Lambda D_{s}$ & 7.20 & 1.05 \\
$\Xi_{c}^{\prime} \Xi  D_{s}$ & 10.80 & 1.95 \\
$\Sigma_{c}  \Sigma D_{s}$ & 7.60 & 0.18\\
$\Xi_{c} \Xi D$ & 3.80 & 0.11\\
$ \Xi_{cc} \Xi_{c} D_{s}$ & 14.60 & 6.03\\
$ \Xi_{cc} \Xi_{c}^{\prime} D_{s}$ &  0.002 & 0.56\\
$ \Omega_{cc} \Omega_{c} D_{s}$ & 29.20  & 13.41 \\
\hline \hline
\end{tabular}
\end{center}


\newpage
\begin{thebibliography}{99}
\bibitem[1] {}  J. K. Bajaj, M. P. Khanna,  and K. Prema, Pramana
{\bf 16}, 249 (1981).
\bibitem[2]  {} M. P. Khanna and R. C. Verma, Z. Phys. C {\bf  47},
275 (1990).
\bibitem[3]   {} S. Coleman and S. L. Glashow, Phys. Rev. {\bf 134
B}, 670 (1964).

\bibitem[4]    {}  R. E. Marshak, Riazuddin,  and C. P. Ryan, Theory
of weak interactions in particle physics, (Wiley Interscience, New
York, 1969) pp. 597.

\bibitem[5] {} M. D. Scadron, Rep. Prog. Phys. { \bf 44}, 213 (1981).

{}.

\bibitem[6] {} J. K. Bajaj, V. S. Kaushal, and M. P. Khanna, Phys.
Rev. {\bf D  18}, 2526  (1978).
\bibitem[7] {} H. Pilkuhn et al., Nucl. Phys. {\bf B 65}, 460
(1973).
\bibitem[8] {} B. Renner and P. Zerwas, Nucl. Phys. {\bf B 35}, 397
(1971).
\bibitem[9] {} M. M. Nagels et al.,  Nucl. Phys. {\bf B 147},  189
(1979).
\bibitem[10] {} Baillon et al., Nucl. Phys. {\bf B 105}, 365 (1976);
{ \it ibid }  {\bf 107}, 189 (1976).
\bibitem[11]  {} H.Y. Cheng and B. Tseng, Phys. Rev. {\bf D 46}, 1042
(1992)    ; T. M. Yan et al., Phys. ReV. { \bf D 46}, 1148 (1992).
\bibitem[12] {} T. Uppal, R.C. Verma, and M.P. Khanna,  Phys. Rev.
{\bf D 49}, 3417 (1994).
\bibitem[13] {} B. W. Lee, C. Quigg and  J. L. Rosner, Phys. Rev.
{\bf D 15}, 157  (1977).

\bibitem[14] {} R. Prasad and  C. P. Singh, Phys. Rev. {\bf D 20},
256 (1979); {\it  ibid } { \bf D 21}, (1980) 836.
\bibitem[15] {} C. B. Dover, S. H. Kahana, and T. C. Trueman, Phys.
Rev. {\bf D 16}, 799 (1977).

\bibitem[16] {}  G. Campbell Jr, Phys. Rev. { \bf D 13}, 662 (1977).

\bibitem[17] {} F. Hussain and K. Khan , Nuo. Cim. {\bf 88 A}, 213
(1985);  S. Pakvasa, S. F. Tuan, and S. P. Rosen, Phys. Rev.{ \bf D
42}, 3746  (1990); R. E. Karlsen and M. D. Scadron, Euro. Phys.
Letts. {\bf 14}, 319 (1991);  Q. P. Xu and A.N. Kamal, Phys. Rev.
{\bf D 46}, 270 (1992).
\end{thebibliography}
\end{document}